%% file: ms.tex
\documentclass{llncs}

\usepackage{stmaryrd}
\usepackage{amsmath}
\usepackage{epstopdf}
\usepackage{amssymb}
\usepackage{pifont}
\usepackage[nocompress]{cite}
\usepackage{caption}
\usepackage{graphicx}
\usepackage{array}
\usepackage{listings}
\usepackage[table]{xcolor}
\usepackage{mathpartir}
\usepackage[vlined,ruled,linesnumbered]{algorithm2e}

\usepackage[vlined,ruled,linesnumbered]{algorithm2e}

\newcommand{\nassimscomment}[1]{
\begin{center}
\fbox{
\begin{minipage}{2.5in}
{\bf Nassim's comment:} {\it #1}
\end{minipage}
}
\end{center}
}

\begin{document}

\title{Data-Flow Guided Slicing}

\author{Mohamed Nassim Seghir}
\institute{University College London}

\definecolor{codegreen}{rgb}{0,0.6,0}
\definecolor{codegray}{rgb}{0.5,0.5,0.5}
\definecolor{codepurple}{rgb}{0.58,0,0.82}
\definecolor{backcolour}{rgb}{0.95,0.95,0.92}
\definecolor{pblue}{rgb}{0.13,0.13,1}
\definecolor{pgreen}{rgb}{0,0.5,0}
\definecolor{pred}{rgb}{0.9,0,0}
\definecolor{pgrey}{rgb}{0.46,0.45,0.48}

 \lstdefinestyle{mystyle2}{
  showspaces=false,
  showtabs=false,
  breaklines=true,
  showstringspaces=false,
  breakatwhitespace=true,
  commentstyle=\color{pgreen},
  keywordstyle=\color{pblue},
  stringstyle=\color{pred},
  basicstyle=\ttfamily,
  moredelim=[il][\textcolor{pgrey}]{},
  moredelim=[is][\textcolor{pgrey}]{}{}
}

 \lstdefinestyle{mystyle}{
    backgroundcolor=\color{backcolour},   
    commentstyle=\color{codegreen},
    keywordstyle=\color{magenta}\bfseries,
    numberstyle=\tiny\color{codegray},
    stringstyle=\color{codepurple}
}

\lstset{
language=Java,
basicstyle=\ttfamily\footnotesize,
numbers=left,
morekeywords={assert, assume},
style=mystyle,
frame=none
}

\newtheorem{Remark}{Remark}[section]
\newtheorem{Example}{Example}[section]
\newcommand{\wuntilop}{\;{\underline{\mathcal{WU}}}\;}
\newcommand{\true}{\mathsf{true}}
\newcommand{\false}{\mathsf{false}}
\newcommand{\program}{P}
\newcommand{\tr}{\pi}
\newcommand{\loc}{\ell}
\newcommand{\stmt}{st}
\newcommand{\location}{\ell}
\newcommand{\statement}{st}
\newcommand{\mysucc}{\mathsf{succ}}

\title{Certificate Enhanced Data-Flow Analysis}
\maketitle

\begin{abstract}
\emph{Proof-carrying-code} was proposed as a solution to ensure a trust relationship between two parties: a (heavyweight) analyzer and a (lightweight) checker. The analyzer verifies the conformance of a given application to a specified property and generates a certificate attesting the validity of the analysis result. It suffices then for the checker just to test the consistency of the proof instead of constructing it. We set out to study the applicability of this technique in the context of data-flow analysis. In particular, we want to know if there is a significant performance difference between the analyzer and the checker. Therefore, we developed a tool, called DCert, implementing an inter-procedural context and flow-sensitive data-flow analyzer and checker for Android. Applying our tool to real-world large applications, we found out that checking can be up to 8 times faster than verification. This important gain in time suggests a potential for equipping applications on app stores with certificates that can be checked on mobile devices which are limited in computation and storage resources. We describe our implementation and report on experimental results. 
\end{abstract}

\input{intro}

\bibliographystyle{abbrv}
\bibliography{biblio}
\end{document}

%% file: intro.tex
\section{Introduction}
Static data-flow analysis has proven its effectiveness in assessing the security of Android applications by identifying data leaks \cite{fuchs:cs-tr-4991, flowdroid, WeiROR14, GordonKPGNR15, BarrosJMVDdE15, ErnstJMDPRKBBHVW14, 0029BBKTARBOM15}. Once we want to install an application on a mobile device, how can we trust the outcome of these tools? The analysis might be broken, or a (malicious) tool can provide a false claim on purpose. Alternatively, it is impractical to directly run the analysis on mobile devices due to scalability issues with static analysis, which is even exacerbated when additional limitations on computing and storage resources are imposed.

\emph{Proof-carrying-code} provides a solution to this problem by ensuring a trust relationship between two parties: a (heavyweight) analyzer and a (lightweight) checker. The analyzer verifies the conformance of a given application to a specified property and generates a certificate attesting the validity of the analysis result. It suffices then for the checker just to test the consistency of the proof instead of constructing it. So far, its applicability has not been studied in the context of data-flow analysis applied to real-world programs. In particular, we want to know if there is a significant performance gain in checking an application compared to analysing it. Hence, we propose a proof-carrying-code inspired data-flow analysis for Android. It has most of commonly desirable features, such as context and flow-sensitivity, conservative handling of aliases and modularity (inter-procedural). In addition, our analysis generates a checkable certificate that is context-independent, thanks to our bottom-up analysis. Hence, a library certificate can be re-used in other applications, provided that the code did not change.

We implemented our approach in a tool called DCert and applied it to real-world large applications. We found out that checking can be up to {\bf 8} times faster than verification. This significant time saving, suggests a new security model for app stores by equipping applications with certificates an deploying lightweight checkers on mobile devices. Our contributions can be summarized in the following:
\begin{itemize}
\item Proposition of certification scheme for data-flow analysis. We are not aware of any other tool that generates a certificate and is scalable to real-world large applications. 
\item Implementation of our approach in a tool called DCert. 
\item Providing empirical evidence of the feasibility of our approach.  
\end{itemize} 
The remainder of the paper is organized as follows: Section~\ref{sec:examples} illustrates our approach via an example. Section~\ref{sec:prelim} and \ref{sec:approach} provide details of our technique and its ingredients. Section~\ref{sec:impl} describes our implementation and reports on experimental results. Finally, Section~\ref{sec:related} surveys related work before concluding with Section~\ref{sec:conclusion}.

\section{Example}
\label{sec:examples}
We start by illustrating our idea through an example. Consider the simple code in Figure~\ref{fig:run_example} as part of an Android application. To ease the presentation, we omit irrelevant details. We have the root procedure \texttt{foo} which makes call to function \texttt{bar} which, in turn, calls procedures \texttt{getId}, \texttt{Send} and \texttt{getNumber}. Function \texttt{getId} reads the device identifier using the API method \texttt{getDeviceId} at line 5. Similarly, function \texttt{getNumber} returns the number of the actual phone making call to the API method \texttt{getLine1Number} at line 5. Finally, procedure \texttt{Send} is used to send the string it takes as argument as an SMS via API method \texttt{sendTextMessage} at line 7. Both methods \texttt{getDeviceId} and \texttt{getLine1Number} represent sources and \texttt{sendTextMessage} is a sink. We want to verify if our app leaks information from certain sources to certain sinks. 
\begin{figure*}[t]
\begin{center}
\begin{tabular}{c}
\begin{tabular}{l@{\hspace{0.4in}}l}
\hline
\\
\begin{minipage}{5cm}	
\begin{lstlisting}[escapechar=\%]
String foo()
{
  String x = bar();
  return x;	
}
\end{lstlisting}
\end{minipage}
    &
\begin{minipage}{6.5cm}
\begin{lstlisting}[escapechar=\%]
String getId()
{      
  // get manager
  TelephonyManager tm = ...; 
  String x = tm.getDeviceId(); 	
  return x;
}

\end{lstlisting}
\end{minipage}

\\

\begin{minipage}{5cm}	
\begin{lstlisting}[escapechar=\%]
String bar()
{
  String x = getId();	
  Send(x);
  String y = getNumber();
  return y;
}

\end{lstlisting}
\end{minipage}
    &
\begin{minipage}{6.5cm}
\begin{lstlisting}[escapechar=\%]
String getNumber(String x)
{
  // get manager		
  TelephonyManager tm = ...; 
  String x = tm.getLine1Number(); 	
  return x;
}

\end{lstlisting}
\end{minipage}


\\
&
\\
\end{tabular}
\\
\begin{minipage}{8.3cm}
\begin{tabular}{c}
\begin{lstlisting}[escapechar=\%]
void Send(String x)
{
  // destination phone number	
  String num = "..."; 
  // get manager
  SmsManager SM = ...; 
  SM.sendTextMessage(num, null, x, ...);
}

\end{lstlisting}
\end{tabular}
\end{minipage}
\\
\\
\hline
\end{tabular}
\end{center}
\caption{Simple Java example illustrating potential data flows from sources to sinks. Method \texttt{getDeviceId} is an Android method for obtaining the device identifier, method \texttt{getLine1Number} permits to obtain the phone number and \texttt{sendTextMessage} allows to send text messages (SMS).}
\label{fig:run_example}
\end{figure*}

\paragraph{\bf Analysis and Certificate Generation.}
Our analysis attempts to find all data leaks, i.e., paths leading from sources to sinks. In addition, it outputs a certificate corroborating its outcome. 

Let us use \texttt{id} and \texttt{num} to respectively refer to the sources \texttt{getDeviceId} and \texttt{getLine1Number}. We also write \texttt{sms} to refer to the sink \texttt{sendTextMessage}. Our analysis computes a summary for each function, which consists of a set of pairs $(x,y)$ expressing the existence of a data-flow from $y$ to $x$. A summary of a given procedure only includes elements visible outside of it. Hence, local variables will not appear in a summary. During the analysis, when a function is invoked from another one, its summary is used instead of re-analysing it. This process is iterated until a fix-point is reached. For illustration, consider Figure~\ref{fig:summar_iter}. It shows the summary computed for the different methods of our previous example (Figure~\ref{fig:run_example}) at each iteration. Initially (iteration 0), all function summaries are empty.   
\begin{figure}
\begin{center}
\begin{tabular}{r @{\hspace{0.1in}}c@{\hspace{0.1in}}|@{\hspace{0.1in}}c@{\hspace{0.1in}}|@{\hspace{0.1in}}c@{\hspace{0.1in}}@{\hspace{0.1in}}|@{\hspace{0.1in}}c}
\cline{2-5}
&	\multicolumn{4}{c}{Iteration} \\
\cline{2-5}
 & {\bf\;\;0\;\;} &{\bf 1} &{\bf 2} & {\bf 3} \\
\cline{2-5}
\
$\mathsf{foo}$: & - & - & - & $\mathsf{(sms, id)}, \mathsf{(ret, num)}$\\
$\mathsf{bar}$: & - & - & $\mathsf{(sms, id)}, \mathsf{(ret, num)}$ & $\mathsf{(sms, id)}, \mathsf{(ret, num)}$\\
$\mathsf{getId}$: & - &$\mathsf{(ret, id)}$ & $\mathsf{(ret, id)}$ & $\mathsf{(ret, id)}$\\
$\mathsf{getNumber}$: & - &$\mathsf{(ret, num)}$ &  $\mathsf{(ret, num)}$ & $\mathsf{(ret, num)}$\\
$\mathsf{Send}$: & - & $\mathsf{(sms, x)}$ & $\mathsf{(sms, x)}$ & $\mathsf{(sms, x)}$ \\
\cline{2-5}
\end{tabular}
\end{center}
\caption{Iterative computation of function summaries. A pair $(x,y)$ models a data-flow from $y$ to $x$.}
\label{fig:summar_iter}
\end{figure}
After iteration 1, empty summaries are still associated with procedures \texttt{foo} and \texttt{bar}, however, summaries for procedures \texttt{getId}, \texttt{getNumber} and \texttt{Send} are updated. The symbol $\mathsf{ret}$ models the return value of a method. Hence, summaries $\mathsf{(ret,id)}$ and $\mathsf{(ret,num)}$, respectively, express flows of the phone identifier (in \texttt{getId}) and the phone number (in \texttt{getNumber}) to a return statement. Similarly, $\mathsf{(sms, x)}$ expresses the presence of a data-flow from the argument $\mathsf{x}$ of procedure $\mathsf{Send}$ to the sink $\mathsf{sms}$. After iteration 2, the summary for procedure $\mathsf{bar}$ is updated as summaries associated with its callees changed in the previous step (iteration 1), meaning potential new data flows. For example, $\mathsf{(sms, id)}$  is due to the path $\mathsf{Send} \leftarrow \mathsf{x} \leftarrow \mathsf{getId}$ in procedure $\mathsf{bar}$, where summaries of procedures $\mathsf{Send}$ and $\mathsf{getId}$ are used. Finally, the last iteration (3) updates the summary for $\mathsf{foo}$ by just propagating $\mathsf{bar}$'s summary. At this state a fix-point is reached and no further changes will be induced. The presence of $\mathsf{(sms, id)}$ in the summary associated with (root) procedure $\mathsf{foo}$ implies a potential leak of the phone identifier via an SMS.

The final map (iteration 3) represents a certificate. It will be returned by the analyser. Our summaries are context-independent, thanks to our bottom-up analysis. Hence, a summary, which also represents a certificate, for a library can be re-used in other applications, provided that the library code is not modified.     

\paragraph{\bf Checking.}
Now the question is how can a client of the analysis trust its claim? The analysis might contain errors or, even worst, an attacker can claim app safety without applying the analysis at all. For this, the computed map will serve as a certificate. To test its validity, we just need to locally check that the summary of each method is valid by assuming the validity of the summaries of its callees. For example, assuming the summary for $\mathsf{bar}$ is $\{\mathsf{(sms, id)}, \mathsf{(ret, num)}\}$, the summary for $\mathsf{foo}$ must be $\{\mathsf{(sms, id)}, \mathsf{(ret, num)}\}$, which is the case. This is performed by the checker, which takes as input a certificate (computed map) and an app, and answers whether the certificate is valid.

Certificate checking is lighter than certificate generation as we do not need to compute a fix-point. Instead, it is performed in a single pass. It has a linear complexity in the number of map entries (functions) and a constant space complexity as we just perform checks without generating information that need to be stored. In what follows, we provide more details about our approach. 

\section{Preliminaries}
\label{sec:prelim}
In this section we provide some ingredients required for the presentation. We start by introducing a language that we consider along our study. 

\subsection{Intermediary Representation}     
Without loss of generality, we consider an object oriented language, accounting for the features distinguishing it from simple imperative languages. We use a Jimple-like \cite{Vallee-RaiGHLPS00} intermediary representation which allows us to encode most of the language constructs using a minimal set of instructions. It has been shown that the whole instruction set of widely deployed virtual machines, such as JVM and Dalvik can be encoded in Jimple \cite{Vallee-RaiGHLPS00,BartelKTM12}. A program is composed of a set of methods, where the body of each method is given by the grammar below:
\[
\begin{array}{rcl}
B &::= &  stmt \\
stmt &::= &  stmt;stmt \;|\; assign \;|\; jmp \;|\; label \;|\; \mathsf{return}\;  id\\
assign &::= & id := c \;|\; id :=  id \;|\; id := \mathsf{op}\;id \;|\;\\
	   &    & id := id \;\mathsf{op}\;id \;|\; id :=  id[id] \;|\; id[id] :=  id  \;|\;\\
	   & &  id :=  id.id \;|\; id.id :=  id \;|\; id := id(id,\ldots,id)\\
jmp &::= &  \mathsf{if}\;cond\;\mathsf{goto}\;label\;|\; \mathsf{goto}\;label\\
cond &::= & id > 0 \;|\; id < 0 \;|\; id = 0\\
\end{array}
\]
While the usual syntax for method invocation in object oriented style is $o.f(\ldots)$, we use a simple form $f(\ldots)$, such that the receiver object is one of the function arguments. The assignment $ret = id$ models a return statement by copying the retuned expression to a special variable $ret$. We use $L$ to denote the language (set of statements) generated by the grammar.      

\subsection{Term Representative}
A key challenge faced when designing any program analysis is aliasing. As we want our analysis to scale to large applications, we need a light but sound solution for the aliasing problem. Therefore, we use the approach proposed by Vall\'ee et al \cite{SundaresanHRVLGG00}. Their idea is based on the observation that only objects of compatible types can be aliases in type-safe programming languages, such as Java. Therefore, a type can be used as a representative for all its object instances. Another advantage of this approach is its simplicity and ease of implementation. 

We introduce the notion of \emph{representative} which is a symbolic representation (over-approximation) of a set of l-value terms that are potential aliases. An l-value is an expression through which a memory location can be updated. In what follows, we provide more details on the notion of representative.

Let $V$ be the set of l-values we can have according to our grammar, i.e, simple variables, array access ($a[i]$) and object field access ($o.f$). We define the function $R: V \longrightarrow V$ which takes an l-value as parameter and returns the corresponding representative. The case where an l-value is a simple term is straightforward. The representative of a simple term $x$ is the term itself as Java does not permit aliasing between simple identifiers. We have 
\[
R(x) = x \;\;\;\text{if $x$ is a simple variable}
\]
The two other cases (field and array access) require more elaboration. 

\subsubsection{Field Access}
Two terms $o_1.f$ and $o_2.f$ are aliases if $o_1$ and $o_2$ points to the same object. Hence, all potential field access that are aliases must map to the same representative. Observe that in type safe languages, like Java, $o_1.f$ and $o_2.f$ can be aliases only if $o1$ and $o2$ belong to the same type hierarchy.   

Given an object $o$ whose declared type is $t$, we write $T(o,f)$ to get the highest class in the type hierarchy of $t$ that contains the field $f$. One way then for over-approximating the representative of a field access $o.f$ is via
\[
R(o.f) = T(o,f).f
\]

Hence $o_1.f$ and $o_2.f$ are mapped to the same representative, provided that the types of $o_1$ and $o_2$ belong to the same type hierarchy.   
\subsubsection{Array Access}
An l-value representing an array access also leads to aliasing. Here the aliasing is due to two causes: reference and index. Two terms $a[i]$ and $a[j]$ refer to the same memory location if $i = j$. In another scenario, terms $a[i]$ and $b[i]$ also refer to the same memory location if $a$ and $b$ points to the same array, i.e, store the same memory reference. We need to take into account both causes of aliasing. To this end, we perform a lightweight alias analysis that split the set of array terms into subsets of potential aliases. Given an array $a$, $A(a)$ returns the unique identifier of the alias set to which $a$ belongs. Note that if we define the representative of $a[i]$ as $A(a)[i]$, it solves the problem of reference-induced aliasing, but does account for the index-induced one. Keeping track of individual indexes, may lead to an infinity of terms when indexes are modified inside a loop. Therefore, we conservatively define the representative of an array access as
\[
R(a[i]) = A(a)
\]    
This takes into account aliases due to index as all element of an array have the same representative.  

\subsection{Control Flow Graph and Call Graph}
Our analysis proceeds at the control flow graph level. Each function is associated with a corresponding control flow graph $\mathit{CFG}$. A $\mathit{CFG}$ is a directed graph $(N,E)$ where $N$ is the set of pairs $(\location,\statement)$, such that $\statement$ is a program basic statement and $\location$ is the corresponding program location. The relation ${E \subseteq N \times N}$
represents the control-flow precedence between program statements. 
We introduce some helper functions to facilitate the presentation. Function $\mysucc(n)$ returns the set of successors of the node taken as argument, $\mathsf{pred}(n)$ returns the set of predecessors of $n$, and $\mathsf{stmt}(n)$ returns the statement associated with the node $n$ in $CFG$. Also $\mathsf{init}(\mathit{CFG})$ returns the initial location (without predecessors) of $\mathit{CFG}$ and $\mathsf{final}(\mathit{CFG})$ returns the final location (without successors).
  
In addition to the control flow graph, representing function intra-information, we need the call graph of the program as our analysis requires inter-information as well.
It is essential that the computed call graph is an over-approximation of the concrete call graph, i.e., any pair of (caller, callee) in real executions of the application is present in the call graph. Java, and object oriented languages in general, have many features, such as method overriding. This makes the construction of an exact call graph (statically) at compile time impossible. Therefore, we over-approximate it using the class hierarchy approach \cite{SundaresanHRVLGG00} which permits to conservatively estimate the runtime types of receiver objects. In what follows, we write $CG(P)$ to denote the call graph of program $P$. It over-approximate the set of all possible pairs of (caller, callee) belonging to the program.  

\section{Data-Flow Certification}
\label{sec:approach}
As seen in section \ref{sec:examples}, our approach has two main components: an analyzer and a checker. The analyzer takes as input an application and produces a certificate. The checker takes as input an application and a certificate, and answers whether the certificate is valid with respect to the input application. As we are in the context of static analysis, a key element shared by both the analyzer and the checker is the abstract domain on which they operate, together with the corresponding transfer function. This is the subject of the next subsection.    

\subsection{Abstract Domain}
A natural way for encoding inter-variable flows is through an abstract domain $D$ representing the powerset of pairs of l-value representatives (section \ref{sec:prelim}). Hence, aliases are taken into account in $D$. We also need to include symbols associated with sources and sinks. Let $SR$ be the set of method identifiers representing sources and $SK$ the set of ones representing sinks. Let us also have a function $b$ which takes a method identifier as argument and returns a corresponding symbol. E.g., as we have seen previously (Section~\ref{sec:examples}), $b(\mathsf{getDeviceId}) = \mathsf{id}$. Function $b$ naturally extends to a set of identifiers $S$ as $b(S) = \{b(x)\;|\; x \in S\}$.
The domain $D$ is then defined as
\[
D  = \mathbb{P}(\{(R(x),R(y)) \;|\; x,y \in (V \cup b(SR \cup SK)) \}),
\]
For consistency, we define $R(x) = x$ if $x \in b(SR \cup SK)$, i.e., the representative of a symbol modeling a source or a sink is the symbol itself.
For a set of pairs in $D$, a pair $(x,y)$ expresses that $y$ flows to $x$.

\subsection{Transfer Function}
To capture variable dependency relations induced by the program, we model the effect of program basic statements, belonging to our language $L$, on elements of the abstract domain $D$. This is achieved through function $F$ that we define below.

Let $L_b$ denotes the set of basic statements in $L$. Given a set of facts $d \in D$ and a statement $s \in L_b$, ${F: D\times L_b \longrightarrow D}$ is formally defined as: 
\begin{align*}
F(d, s) = & \{ (x,y) \;|\; \exists z.\; (x,z) \in \mathsf{flow}(s) \wedge (z,y) \in d\}  
\cup (d - \mathsf{kill}(s, d)).
\end{align*}

In the formula above, function $\mathsf{flow}$ returns the set of flows that are locally induced by the statement taken as argument. 
For example $\mathsf{flow}(x:= y + z)$ yields $\{(x,y), (x,z)\}$. Function $F$ transitively extends the relation represented by the input facts $d$ in combination with the relation induced by the input statement $s$. It also uses function $\mathsf{kill}$ to exclude facts that are no longer valid after the assignment. For example
\[
F(\{(x,t), (y,p)\}, x:= y + z) = \{(x,p), (y,p)\}.
\]          
As the assignment modifies variable $x$, the fact $(x,t)$ no longer holds. The fact $(x,p)$ is obtained by transitivity from the input fact $(y,p)$ in combination with $(x,y)$ which is induced by the assignment statement. 
We provide definitions of functions $\mathsf{flow}$ and $\mathsf{kill}$ for the different kind of basic statements in Table~\ref{tab:dep_kill}.

Assignments to simple variables, representing the first group in Table~\ref{tab:dep_kill}, generate pairs expressing that identifiers appearing in the assignment right-hand-side flow to its left-hand-side. For the cases $id_1 := id_2.id_3$ and $id_1 := id_2[id_3]$, we use the representative of the assignment left-hand-side to take aliases into account. If the right-hand-side is a constant, no flows are generated. The return statement $\mathsf{return}\;id$ is handled as a simple assignment $ret := id$, where $ret$ is a special variable. In all these cases, we kill input facts expressing previous flows to the assignment left-hand-side. 

In case of an assignment to an object field or array element (second group), we use the representative of the assignment left-hand-side to take aliases into account. However, we do not kill any fact to preserve soundness. Indeed, a representative is an over-approximation of possible aliases. Therefore, the updated l-value may or may not be an actual alias of a given fact.

For a function call (third group), we first extract its summary ($\mathsf{summary}[f]$), which is the set of facts expressing the flows induced by the function over program variables. We then replace formal parameters of the function with the corresponding actual ones in each fact. We also replace the special variable $ret$ with the actual return-to variable $r$. Note that summaries for functions are computed iteratively, on-the-fly, during the analysis as we will see later.

Finally, a conditional $cond$ does not have any effect on the input facts, making the transfer function $F$ behave as an identity function. 
\begin{table*}[t]
\begin{center}
\begin{tabular}{lcc}
\hline
 Statement: $\stmt$ & $\mathsf{flow}(\stmt)$ & $\mathsf{kill}(\stmt, d)$\\
\hline
\hline
 $id := c$ & $\emptyset$ & $\{(x,y) \in d \;|\; x = id\}$\\

 $id_1 := id_2$ & $\{(id_1, id_2)\}$& $\{(x,y) \in d \;|\; x = id_1\}$\\

$id_1 := \mathsf{op}\;id_2$ & $\{(id_1,id_2)\}$&  $\{(x,y) \in d \;|\; x = id_1\}$\\

$id_1 := id_2 \;\mathsf{op}\;id_3$ & $\{(id_1, id_2), (id_1, id_3)\}$& $\{(x,y) \in d \;|\; x = id_1\}$\\
 $id_1 := id_2.id_3$ & $\{(id_1, R(id_2.id_3))\}$&  $\{(x,y) \in d \;|\; x = id_1\}$\\

 $id_1 := id_2[id_3]$ & $\{(id_1, R(id_2[id_3]))\}$&  $\{(x,y) \in d \;|\; x = id_1\}$\\
 $\mathsf{return}\;id$ & $\{(ret,id)\}$ & $ \emptyset$\\
 
\hline
 $id_1.id_2 := id_3$ & $\{(R(id_1.id_2), id_3)\}$&  $\emptyset$\\

 $id_1[id_2] := id_3$ & $\{(R(id_1[id_2]), id_3)\}$&  $\emptyset$\\
\hline

$r = f(id_1,\ldots,id_n)$ & $s[id_1/x_1\ldots,id_n/x_n, r/ret]$& $d - \{(x,y) \in d \;|\; x = r\}$\\
& ($s = \mathsf{summary[f]}$ &\\
& and $x_1,\ldots,x_n$ are formal parameters of $f$) &\\
\hline
 $cond$ & 
$\emptyset$ &  $\emptyset$\\
\hline
\end{tabular}
\end{center}
\caption{ Definition of functions $\mathsf{flow}$
 and $\mathsf{kill}$ for the different kind of basic statements considered in our language.
}
\label{tab:dep_kill}
\end{table*}
In the next subsections, we present the core components of our approach: analyser and checker.

\subsection{The analyser}
In what follows, we describe the algorithm that performs the data-flow analysis. A key feature of our algorithm is that it produces a certificate. 

To compute all possible flows from sources to sinks, we propose a bottom-up inter-procedural data-flow analysis that computes method summaries that are context-independent. A summary is an over-approximation of the relation entailed by a given method over l-values used in the program.    
\begin{algorithm}[h!]
\KwIn {Program $P$}
\KwOut {map from methods to sets of facts (certificate)}
\SetKw{Var}{Var}
\SetKw{continue}{continue}
\Var map $summary$\;
\Var list $\mathit{WL}$\;

$\mathit{WL}$ := $\{\text{methods of }P\}$\;

\ForEach{$ m \in \mathit{WL} $}
  	{
  	  $summary[m] = \{\}$\;
	}	

\While{$\mathit{WL} \neq \emptyset$}
{
  $m$ := $\mathsf{Top}(\mathit{WL})$\;
  $\mathsf{Pop}(\mathit{WL})$\; 	
  $s_{new}$ := $\mathsf{Summarise}(m)$\;
  $s_{new}$ := $\{(x,y) \in s_{new} \;|\; x,y \not \in \mathsf{local(m)}\}$\;  
  \If{$s_{new} \neq summary[m]$}
     {
     	$summary[m]$ := $s_{new}$\;
     	\ForEach{$(m',m) \in CG(P)$}
			{
			 	$\mathsf{Add(\mathit{WL}, m')}$\;
			}	
     }
        
}	
\Return $summary$\;
\caption{Analyzer}
\label{alg:analyser}
\end{algorithm}
Our inter-procedural analysis is implemented via algorithm \textsf{Analyser} (Algorithm~\ref{alg:analyser}), which takes as input a program $P$ and produces a certificate that consists of the summary map.  First, all summary entries are initialised (line 5). Second, a work-list based procedure is applied to compute a fix-point (lines 6-14). Summaries are computed calling procedure \textsf{Summarise} (line 9). At line 10, we discard pairs in which at least one element is local to the current method $m$. Function \textsf{local} returns the set of representatives corresponding to the local variables of the method taken as argument. It is useless to keep elements referring to local variables in a summary as they are invisible outside the method. 

Summaries are updated until no new changes occur. This check is carried out at line 11. 
If the summary of a method is updated then all its callers, need to be analyzed again (lines 13 and 14). Recall that $CG(P)$ corresponds to the call graph of program $P$ (section \ref{sec:prelim}). Hence, $(m',m)$ is a pair of caller and callee. Finally, the summary map, which also represents a certificate, is returned. 

More concretely speaking, algorithm \textsf{Analyser}, implements two fix-point iterations which are invoked hierarchically. The inner iteration is applied at the method level by calling algorithm \textsf{Summarise} (Algorithm~\ref{alg:summarise}). Its role is to compute an approximation of the flow relations over program variables induced by the method $m$ taken as argument. It proceeds by computing the transitive closure 
over elements from the domain $D$ with respect to statements of $m$, applying the transfer function $F$. We implement it as a standard iterative work-list procedure. Fist, all locations are initialised with empty sets (lines 6 and 7) apart from the initial location (line 8) with which the set of pairs of l-value representatives ($V$), in addition to sources and sinks, is associated. A fix-point computation is then carried out (lines 10-18) where new facts are produced by simulating the affect of program statements on input facts (line 15) . When the set of facts associated with a location is updated, all successor locations need to be considered (lines 17-18). Once a fix-point is reached, the algorithm returns the set of facts accumulated at the final location (line 19). 

In the next section we describe the algorithm implementing the checker.
\begin{algorithm}[h!]
\KwIn {method $M$}
\KwOut {set of facts}
\SetKw{Var}{Var}
\Var list $\mathit{WL}$\;
\SetKw{continue}{continue}
\Var map $\mathit{IN}$, $\mathit{OUT}$\;
Let $\mathit{CFG}$ be the control flow graph of $m$\;
Let $\mathit{V_M}$ be the set of l-values appearing in $M$\;
\ForEach{$n \in \mathsf{node}(\mathit{CFG})$}
  	{
  		$\mathit{IN}[n] = \{\}$\;
  		$\mathit{OUT}[n] = \{\}$\;	
	}	
$\mathit{IN}[\mathsf{init}(\mathit{CFG})] = \{(R(x),R(x))\;|\; x \in (\mathit{V} \cup              
b(SR \cup SK))\}$\;

$\mathit{WL}$ := $\{\mathsf{init}(CFG)\}$\;
\While{$\mathit{WL} \neq \emptyset$}
{
  $n$ := $\mathsf{Top}(\mathit{WL})$\;
  $\mathsf{Pop}(\mathit{WL})$\; 	
  $\mathit{OUT_0}$ := $\mathit{OUT}[n]$\;
  $\mathit{IN}[n]$ := $\bigcup \mathit{OUT}[n'] \;\;\text{such that}\; n' \in \mathsf{pred}(n)$\;
  $\mathit{OUT}[n]$ := $F(\mathit{IN}[n],\mathsf{stmt}(n))$\;
  \If{$\mathit{OUT_0} \neq \mathit{OUT}[n]$}
     {
     	\ForEach{$n' \in \mathsf{succ}(n)$}
			{
			 	$\mathsf{Push}(\mathit{WL}, n')$\;
			}	
     }
        
}	
\Return $\mathit{OUT}[\mathsf{final}(\mathit{CFG})]$\;
\caption{Summarise}
\label{alg:summarise}
\end{algorithm}

\subsection{The checker}
The checker component takes as input a program and certificate, and answers whether the certificate is valid with respect to the input program. This is implemented via procedure \textsf{Checker} (Algorithm \ref{alg:checker}). 
First, it extracts all methods in the program (line 2). For each method it checks its presence in the certificate, which corresponds to the first disjunct of the test at line 4. If this is not the case then the certificate is invalid. Indeed, all methods of the program must have entries in the certificate. This prevents circumventing the checker by only providing entries for safe methods in the certificate. The second disjunct in the test (line 4) checks if the actual computed summary corresponds to the one given by the certificate. 
As we can see, certificate checking is linear in the number of methods as it is done in a single pass (lines 3-5). Moreover, it has a constant memory complexity as no additional information needs to be stored.     

\begin{algorithm}[h!]
\KwIn {Program $P$, Map (certificate) $C$}
\KwOut {Boolean}
\SetKw{Var}{Var}
\SetKw{continue}{continue}
\Var list $\mathit{L}$\;
$\mathit{L}$ := $\{\text{methods of }P\}$\;
\ForEach{$ \text{method } m \in L $}
  	{
          \If{$m \not \in C \vee \mathsf{Summarise}(m) \neq C[m]$}
             {
               \Return $\false$\;
             }  	  
	}
\Return $\true$\;
\caption{Checker}
\label{alg:checker}
\end{algorithm}

\subsection{Discussion} 
We discuss some aspects related to our analysis without deep diving into detail.s
\paragraph{\bf Implicit Dependencies.}
The kind of dependencies treated in the previous section are due to assignments. There exist indirect dependencies resulting from conditional statements. Consider the following examples 
\begin{center}
\begin{verbatim}
                  y = 0; if (x > 0) {y = z;}.
\end{verbatim}
\end{center}
If we just consider direct dependencies, we conclude that variable \texttt{y} only depends on \texttt{z}. However, depending on whether variable \texttt{x} is positive or not, the value of \texttt{y} may vary as the assignment inside the conditional may or may not be executed. We say that there is a control or implicit dependency between \texttt{x} and \texttt{y}. Hence, a sound approximation of dependencies of \texttt{y} must include both \texttt{x} and \texttt{z}. We use a well-established approach proposed by Ferrante et al~\cite{FerranteOW87} to compute control dependencies.
\paragraph{\bf Handling Unavailable Code.}
One challenge we have faced is taking into account library calls. As the code is often unavailable, we need to over-approximate the effect of library APIs on program variables. Our solution is similar to the one adopted by Flowdroid \cite{flowdroid}: we use two rules to model the effect of library calls. The first rule assumes that the result of a method depends on its parameters as well as the receiver object. The second rule assumes that the receiver object depends on the method parameters. For example, for a method that appends a character to a string, we have a rule modelling that the result depends on the appended character.     

\paragraph{\bf Reflection.}
Reflection is a common obstacle to static analysis as it obfuscates destinations of method invocations. It is impractical to naively consider all methods to which a potential call site may resolve. Some solutions were proposed in the literature to deal with reflection \cite{BarrosJMVDdE15}. We plan to investigate the combination of such approaches with our technique.
   
\section{Implementation and Experiments}
\label{sec:impl}
We have implemented our approach in a tool called DCert, which is written in Python and uses Androguard\footnote{https://github.com/androguard} as front-end for parsing and decompiling Android applications. It accepts Android applications in bytecode format (APK), so no source code is required. One can simply download an app from a store of choice and analyze it. As mentioned previously, DCert has two main components: Analyzer and Checker.   

The analyzer takes as input an app and outputs an analysis report (flows found) together with a certificate. The checker takes an application and a certificate as input, and answers whether the certificate is valid with respect to the application taken as input. 

We performed experiments on 13 real-world popular applications, from the Google Play store\footnote{https://play.google.com/store/apps}, ranging over different domains: communication, office, social, etc. We use a typical Linux desktop in our experiments. 

First, we apply the analyzer to an application and obtain a certificate. Then, we invoke the checker on the same application in addition to providing the previously generated certificate as input. Results are illustrated in Table~\ref{tab:experiments}. Column \#Methods shows the number of methods per application as an indicator of the application size. The two next columns provide the analysis and checking time. They show times purely taken by the analysis and checking process. We do not consider pre-processing, such as decompilation, etc. Column \#Leaks provide the number of data leaks from sources to sinks found by our analysis. Due to application size and non-availability of source code, we did not investigate if the current leaks represent a ground truth. 

Results show a significant difference between the analysis and checking time in most of the cases. Checking can sometimes be more than {\bf 8} times faster as is the case for the {\bf WhatsApp} application.  

\begin{figure}[!ht]
\begin{center}
\includegraphics[scale=0.5]{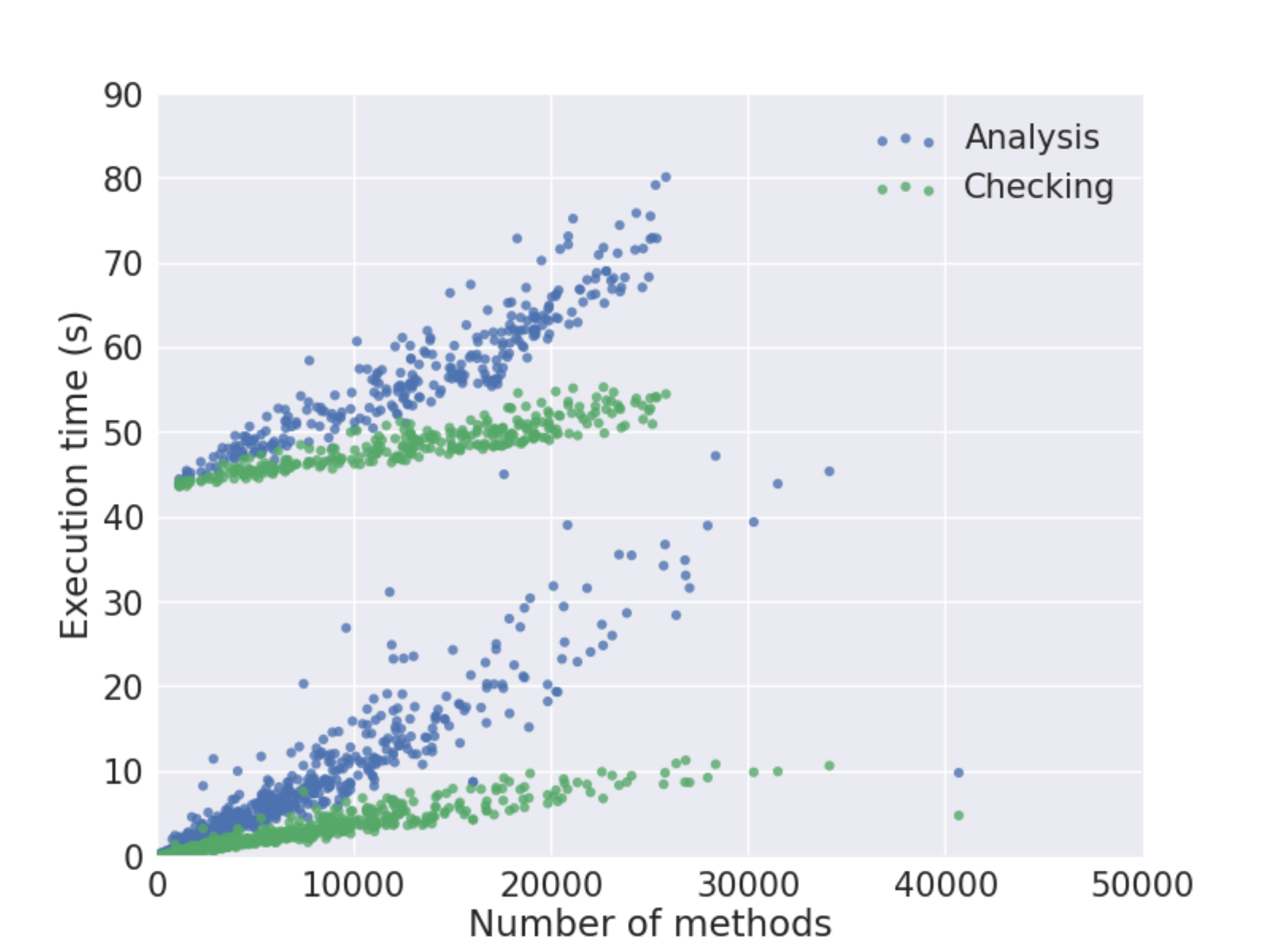}
\end{center}
\caption{Analysis (in blue) and checking (in green) times in function of method number per application. Results for 1070 applications are included.}
\label{fig:fig_1} 
\vspace{-.5cm}
\end{figure}

As we think that there is a correlation between the application size (number of methods) and the gap between analysis and checking time, we further investigate this hypothesis by applying our approach to 1070 apps randomly collected from Androzoo\footnote{https://androzoo.uni.lu/}. Androzoo apps have various origins, including the Google Play store which is the predominant source of most of the apps. Results are illustrated in Figure~\ref{fig:fig_1}, which shows the analysis and checking time per application in function of the number of methods. Dots in blue represent analysis time and the ones in green correspond to checking time. We have two clusters, and in each one, we clearly see that analysis time is higher than checking time. Moreover, the gap between the two increases as application size gets larger. This confirms our hypothesis and further supports our certification scheme proposal, as applications tend to be larger and larger with new features added, backward compatibly guarantees, etc.
\begin{table}[h]
\begin{center}
\begin{tabular}{c|c@{\hspace{0.2in}}c@{\hspace{0.2in}}c@{\hspace{0.2in}}c}
\hline
App & \# Methods & Analysis time (s) & Checking time (s) & \# Leaks \\
\hline
Instagram  &  50754  &  386.37  &  65.75  &  2 \\
Skype  &  42269  &  59.73  &  19.05  &  0 \\
Firefox  &  43076  &  83.53  &  17.9  &  4 \\
Uber  &  44875  &  125.75  &  67.69  &  4 \\
Messenger  &  6139  &  11.2  &  3.73  &  0 \\
Google\_Chrome  &  39182  &  101.19  &  24.68  &  2 \\
Facebook  &  6012  &  11.17  &  3.58  &  0 \\                                                                                                         
Google\_Maps  &  37270  &  478.98  &  69.65  &  0 \\                                                                                                   
WhatsApp  &  47156  &  800.63  &  96.74  &  13 \\                                                                                                     
Acrobat\_Reader  &  40801  &  89.55  &  16.75  &  4 \\                                                                                                 
YouTube  &  52480  &  383.38  &  99.17  &  2 \\
MicrosoftWord  &  47752  &  92.0  &  23.35  &  0 \\
Viber  &  45232  &  94.22  &  19.68  &  1 \\
\hline
\end{tabular}
\end{center}
\caption{Analysis and Checking results for 13 popular apps from the Google Play store.} 
\label{tab:experiments}
\end{table}

\section{Related Work} 
\label{sec:related}
Our work is related to many topics: certification, taint analysis and Android security. Along these axes we report on related work.   

The idea of associating proofs with code was initially proposed by Necula under the moniker \emph{Proof-Carrying Code} (PCC) \cite{NeculaL96, Necula97}. It was then used to support resource policies for mobile code~\cite{AspinallM05,BartheCGJP07}. Furthermore, Desmet et al. presented an implementation of PCC for the .NET platform \cite{DesmetJMPPSV08}. While the tool EviCheck \cite{SeghirA15, Seghir0M16} is also based on a similar idea and targets Android, it is unable to analyse data-flow properties. Cassandra also applies PCC to Android \cite{LortzMSBSW14}. Their approach proposes a type system to precisely track information flows. While precision is an advantage, it is hard to assess the practicability of their approach as no experiments involving real-world applications are reported. Our approach is applicable to real-world large applications.      

\emph{Taint analysis} is a technique used in software security \cite{flowdroid, Enck:2014, WeiROR14, TrippPCCG13, GordonKPGNR15} to find data leaks from some given sources to some given sinks. Our Analysis generates a certificate which is not the case for the mentioned approaches. In addition, thanks to our bottom-up inter-procedural, summaries computed via our technique are context-independent. This allows verification re-use. 

Android security is an active area of investigation, many tools for analyzing security aspects of Android have emerged. Some rely on dynamic analysis \cite{aurasium, EnckGCCJMS10,reina:copperdroid,ZhangYXYGNWZ13,BackesGHMS13}. Other tools are based on static analysis \cite{flowdroid, scandal, FahlHMSBF12,AuZHL12,ChinFGW11, ChenJDDMMWRS13, JeonMVFRFM12}. We are interested in the last category (static analysis) as our aim is to certify the absence of bad behaviors. Our work is a complement to these tools as we are not only interested in analyzing applications, but also to return a verifiable certificate attesting the validity of the analysis result.

\section{Conclusion and Further Work}
\label{sec:conclusion}
We have presented a data-flow certification approach inspired by the proof-carrying-code idea. It consists of splitting the verification process between a heavyweight analyzer and a lightweight checker and using a certificate to ensure trust between the two elements. We implemented our technique in a tool called DCert and tested it on real-world applications. Results show that checking can be up to 8 times faster than verification. Our finding suggests a potential for deploying the checker on mobile devices which are limited in resources. This will be the subject of our future investigations. We are not aware of any other tool that implements a similar certification scheme and is scalable to real-world large applications.